\documentclass[preprint, showpacs]{revtex4}
\usepackage[english]{babel}
\usepackage{graphicx, epsfig}
\usepackage{amscd}
\usepackage{amsmath,amsfonts,amssymb}

\begin{document}
\title{Quantum correlations in a system of nuclear $s=1/2$ spins in a strong magnetic field}

\author{E.\,B.\,Fel'dman} 
\author{E.\,I.\,Kuznetsova}
\email{kuznets@icp.ac.ru}
\author{M. A. Yurishchev}

\affiliation{Institute of Problems of Chemical Physics of the Russian Academy of Sciences, 142432
Chernogolovka, Moscow Region, Russia}


\begin{abstract}{Entanglement and quantum discord for a pair of nuclear spins $s=1/2$ in a nanopore filled with a gas of spin-carrying molecules (atoms) are studied. The correlation functions describing dynamics of dipolar coupled spins in a nanopore are found. The dependence of spin-pair entanglement on the temperature and the number of spins is obtained from the reduced density matrix, which is centrosymmetric (CS). An analytic expression for the concurrence is obtained for an arbitrary CS density matrix. It is shown that the quantum discord as a measure of quantum correlations attains a significant value at low temperatures. It is shown also that the discord in the considered model has ``flickering'' character and disappears periodically in the course of the time evolution of the system. The geometric discord is studied for arbitrary $4\times 4$ CS density matrices.}
\end{abstract}



\maketitle

\section{{\small Introduction}}

Quantum correlations in many-particle systems are responsible for the performance of the quantum devices (in particular, quantum computers) and give them significant advantages over their classical counterparts ~\cite{vedral_arxiv, uh}. Until recently it was believed \cite{fazio}  that entanglement is responsible for quantum correlations \cite{braunstein}, and quantum devices can be created only on the basis of  materials which can be easily prepared in entangled states \cite{uh}. However, it turned out that one can work out quantum algorithms ~\cite{braunstein, lanyon} which significantly outperform their classical counterparts by mixing separable (non-entangled) states. It also turned out that quantum non-locality can be observed in systems without entanglement \cite{bennett}. From this, one can conclude that entanglement describes only a part of quantum correlations but not all of them.

According to the current understanding \cite{vedral_arxiv}, total correlations, quantum and classical, are determined by the mutual information \cite{vedral2}. The problem is how to divide the correlations into quantum and classical ones. The problem was solved independently by Henderson and Vedral ~\cite{henderson} as well as independently Ollivier and Zurek ~\cite{olliver}. The classical correlation in a binary system is determined by the total set of projective measurements carried out only for its one subsystem ~\cite{henderson}. Then the quantum correlation measure (the quantum discord) is determined as the difference between the mutual information and its classical part optimized over all possible projective measurements \cite{henderson, olliver}. The quantum discord is determined completely by quantum properties of the system and equals zero for classical systems.

A calculation of the quantum discord is a complex problem because it is necessary to execute many-parametric optimization. Therefore, analytic calculations for the discord are possible only in simple two-qubit systems \cite{luo, ali}. 

A special attention was devoted to a connection of the quantum discord and physical parameters of the system \cite{feldman, yur}. Such investigations open a direct way to experimental measurements of the discord. In the present article we investigate theoretically the entanglement and quantum discord in a thin silicon film containing close nanopores filled with a a gas of spin-carrying molecules (atoms) with spin  $s=1/2$ in a strong external magnetic field. NMR line shape in such films was studied experimentally \cite{baugh} and theoretically \cite{rud, fedorova}. Fast molecular motion in nanopores do not average the dipole-dipole interactions (DDI) completely (space confinement) and the residual DDI is described by one coupling constant which is the same for all pairs of interacting spins \cite{baugh, rud}. Reducing the density matrix (which describes the time evolution of the system with the residual DDI) over all spins except the chosen pair one can obtain information about the pair entanglement and quantum discord in the system. The main goal of this article is the investigation of entanglement and quantum discord of spin pair in the nanopore filled with a gas of spin-carrying molecules (atoms).

It is also significant that the reduced density matrix in the considered model is centrosymmetric (CS) \cite{centrosymmetric}. In the present article we obtain an analytic expression for the concurrence \cite{wootters} of an arbitrary $4\times 4$ CS density matrix. The geometric discord for an arbitrary $4\times 4$ CS-matrix is also calculated.

\section{{\small The reduced density matrix of the spin pair in a nanopore.}}

\begin{figure}
\includegraphics[scale=0.37]{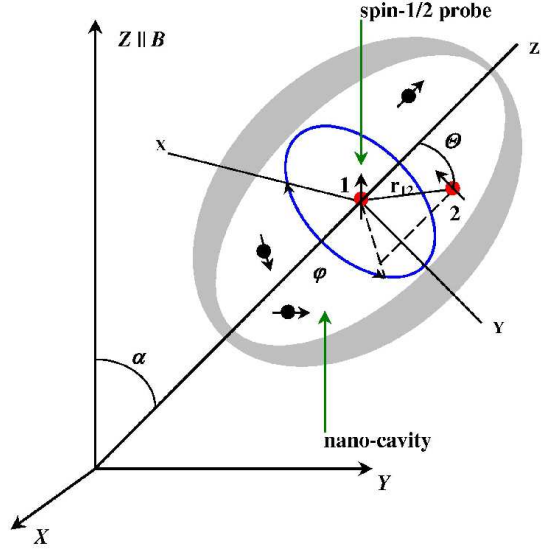}
\caption{A gas of spin-carrying molecules (atoms) in a nanopore in the external magnetic field.}
\label{pic:graf1}
\end{figure}

We consider a system consisting of $N$ spin-carrying molecules (atoms) of a gas with spin $s=1/2$ in a closed nanopore in a strong external magnetic field $B$ (Fig. \ref{pic:graf1}). Molecular motions lead to a partial averaging of the secular DDI, $H_{dz}$ \cite{baugh, rud}, and the Hamiltonian of the residual (averaged) DDI is \cite{rud}
\begin{equation}
H_{dz}=\frac{D}{2}(3I_z^2-I^2),
\end{equation}
where $D$ is the coupling constant, which is the same for all spin pairs \cite{baugh, rud}, $I^2$ is the square of the total angular momentum, $I_{\alpha}= \sum_{i=1}^N I_i^{\alpha}$, $I_i^{\alpha}$ is the operator of the momentum projection of spin $i$ ($i=1,2,...,N$) on the axis $\alpha$ ($\alpha=x,y,z$). In the initial moment of time the resonance $\pi/2$-pulse,  turning spins by the angle $\pi/2$ about the axis $y$ (Fig. \ref{pic:graf1}), acts on the system. As a result, the equilibrium density matrix, $\rho_0$, in a strong external magnetic field is given by 
\begin{equation}
\rho_0=\frac{1}{Z}e^{\beta I_x},
\end{equation}
where $Z=\texttt{Tr}\{e^{\beta I_x}\}$ $=2^N \cosh ^{N}\frac{\beta}{2}$ is the partition function, $\beta=\frac{\hbar\omega_0}{k_B T}$ is the inverse dimensionless temperature, $\omega_0=\gamma B_0$ ($\gamma$ is the gyromagnetic ratio) is the Larmour frequency, and $T$ is the temperature of the system. The time evolution of the density matrix is described as
\begin{equation}\label{matr}
\rho(t)=e^{-iH_{dz}t}\rho_0 e^{iH_{dz}t}=\frac{1}{Z}e^{-iatI_z^2}e^{\beta I_x}e^{iatI_z^2},
\end{equation}
where we took into account that $[I_\alpha, I^2]=0\; (\alpha=x,y,z)$ and set $a=\frac{3D}{2}$. The density matrix \eqref{matr} describes the free induction decay and NMR line shape \cite{rud, fedorova}. In order to solve our problems it is convenient to rewrite the density matrix $\rho(t)$ as \cite{doronin}
\begin{equation}
\rho(t)=\sum_{\xi_1,\xi_2,..., \xi_N=0}^3 \alpha_{1,2,...,N}^{\xi_1,\xi_2,..., \xi_N}(t)I_1^{\xi_1}\otimes I_2^{\xi_2}\otimes ...\otimes I_N^{\xi_N},
\end{equation}
where $\xi_j=0,1,2,3\; (j=1,2,...,N),\; I_j^0=E_j$ 
 ($E_j$ is the $2\times 2$ matrix ), $I_j^1=I_j^x, I_j^2=I_j^y$, $I_j^3=I_j^z$, and $\alpha_{1,2,...,N}^{\xi_1,\xi_2,..., \xi_N}(t)$ is a function of time. Since $\texttt{Tr}(I_j^{\xi_j})=0$ $(j=1,2,...,N, \xi_j=1,2,3)$, the condition of the normalization of the density matrix leads to
\begin{equation}\label{alph00}
\alpha_{1,2,...,N}^{0,0,..., 0}=\frac{1}{2^N}.
\end{equation}

Consider an arbitrary spin pair. Without loss of generality we can assume that it includes spins 1 and 2. We will investigate the entanglement and the quantum discord for that pair. The reduced over spins 3, 4, ..., N density matrix $\rho^{(1,2)}(t)$ is
\begin{equation}
\rho^{(1,2)}(t)=\sum_{\xi_1,\xi_2=0}^3 \alpha_{1,2}^{\xi_1,\xi_2}(t)I_1^{\xi_1}\otimes I_2^{\xi_2},
\end{equation}
where $\alpha_{1,2}^{\xi_1, \xi_2}(t)=\alpha_{1,2,3...,N}^{\xi_1,\xi_2, 0,...,0}(t)2^{N-2}$. One can verify that $\texttt{Tr}\{\rho^{(1,2)}\}=1$. Functions $\alpha_{1,2}^{\xi_1, \xi_2}(t)$ can be expressed via the correlation functions of the spin system as follows:
\begin{equation}\label{alph}
\alpha_{1,2}^{\xi_1, \xi_2}(t)=
\left\{\begin{array} {l}
4 \texttt{Tr}\{\rho(t)(I_1^{\xi_1}\otimes I_2^{\xi_2}\otimes E_3\otimes...\otimes E_N)\},\\
\hspace{1cm} \; \xi_1, \xi_2=1,2,3,\\
  \texttt{Tr}\{\rho(t)(I_1^{\xi_1}\otimes I_2^{\xi_2}\otimes E_3\otimes...\otimes E_N)\},\\
   \; \xi_1=0, \xi_2=1,2,3 \;\mbox{or}\;  \xi_1=1,2,3, \xi_2=0,\\
\frac{1}{4}\texttt{Tr}\{\rho(t)(I_1^{\xi_1}\otimes I_2^{\xi_2}\otimes E_3\otimes...\otimes E_N)\}, \\
\hspace{1cm} \xi_1=0, \xi_2=0.
\end{array}\right.
\end{equation}
One can find from Eq. \eqref{alph00} that $\alpha_{12}^{00}=1/4$. The relationships \eqref{matr}, \eqref{alph} lead to
\begin{equation}\label{ustanovl}
\alpha_{12}^{03}=\alpha_{12}^{30}=\alpha_{12}^{12}=\alpha_{12}^{21}=\alpha_{12}^{31}=\alpha_{12}^{13}=0.
\end{equation}
Taking into account the permutation symmetry over the numbers of spins 1 an 2, it is convenient to introduce the correlation functions
\begin{eqnarray}\label{Qf1}
&&
p=\texttt{Tr}\{\rho(t)(I_1^{x}\otimes E_2\otimes E_3\otimes...\otimes E_N)\}\\\nonumber && =\texttt{Tr}\{\rho(t)(I_2^{x}\otimes E_1\otimes E_3\otimes...\otimes E_N)\},\\
\label{Qf2} &&
q=\texttt{Tr}\{\rho(t)(I_1^{x}\otimes I_2^{x}\otimes E_3\otimes...\otimes E_N)\},\\
\label{Qf3} &&
r=\texttt{Tr}\{\rho(t)(I_1^{y}\otimes I_2^{y}\otimes E_3\otimes...\otimes E_N)\},\\
\label{Qf4} &&
u=\texttt{Tr}\{\rho(t)(I_1^{z}\otimes I_2^{y}\otimes E_3\otimes...\otimes E_N)\}\\ \nonumber
 &&
=\texttt{Tr}\{\rho(t)(I_1^{y}\otimes I_2^{z}\otimes E_3\otimes...\otimes E_N)\},\\
\label{Qf5} &&
v=\texttt{Tr}\{\rho(t)(I_1^{z}\otimes I_2^{z}\otimes E_3\otimes...\otimes E_N)\}.
\end{eqnarray}
Using Eq. \eqref{matr} for the density matrix $\rho(t)$ and executing the unitary transformation $U_1=\exp(i\pi I_1^x)\otimes E_2\otimes E_3 \otimes...\otimes E_N$ under the sign of the trace \eqref{Qf5} we obtain that $v=0$. Using Eqs. \eqref{alph}-\eqref{Qf5} the reduced density matrix can be written as
\begin{equation}\label{redmatr}
\rho^{(1,2)}(t)=\left( \begin{matrix} \frac{1}{4} & \frac{p}{2}-iu & \frac{p}{2}-iu & q-r\\
\frac{p}{2}+iu & \frac{1}{4}& q+r& \frac{p}{2}+iu\\
\frac{p}{2}+iu & q+r& \frac{1}{4}& \frac{p}{2}+iu\\
q-r & \frac{p}{2}-iu & \frac{p}{2}-iu & \frac{1}{4} \end{matrix} \right)
\end{equation}

It is obvious that the density matrix elements $\rho^{(1,2)}_{i,j}(t)$ $(i,j=1,2,3,4)$  satisfy to the relationship $\rho^{(1,2)}_{i,j}(t)=\rho^{(1,2)}_{5-i,5-j}(t)$. Such matrices are called centrosymmetric \cite{centrosymmetric}. One can show that the general expression for the hermitian CS $4\times 4$ matrix is
\begin{equation}\label{csmatrix}
\rho=\left( \begin{matrix} p_1 & p_2+ip_3 & p_4+ip_5 & p_6 \\
p_2-ip_3 & \frac{1}{2}-p_1& p_7 & p_4-ip_5\\
p_4-ip_5 & p_7 & \frac{1}{2}-p_1& p_2-ip_3\\
p_6 & p_4+ip_5 & p_2+ip_3 & p_1 \end{matrix} \right)
\end{equation}
where $p_i\, (i=1,...,7)$ are real parameters. Because the density matrix must be nonnegative defined, all its eigenvalues are greater or equal to zero:
\begin{equation}\label{cseiginv}
\begin{array}{l}
\Lambda_{1,2}=\frac{1}{2} (p_6+p_7+1/2)\pm\sqrt{\frac{1}{4}(2p_1+p_6-p_7-1/2)^2+(p_2+p_4)^2+(p_3+p_5)^2}\geqslant 0, \\
\Lambda_{3,4}=\frac{1}{2} (1/2-p_6-p_7)\pm\sqrt{\frac{1}{4}(2p_1-p_6+p_7-1/2)^2+(p_2-p_4)^2+(p_3-p_5)^2}\geqslant 0.
\end{array}
\end{equation}
These inequalities restrict the range of values for the parameters $p_1,... p_7$.

\section{{\small Calculation of the correlation functions}}

To calculate of the correlation functions \eqref{Qf1}-\eqref{Qf5} we use the commutation relations for spin 1/2 \cite{landau} and the formula \cite{rud}
\begin{equation}
e^{-iatI_z^2}I^+e^{iatI_z^2}=e^{-iat(2I_z-1)}I^+,
\end{equation}
where $I^+=I_x+iI_y$. As a result, the expressions under the trace sign in Eqs. \eqref{alph}-\eqref{Qf4} can be represented as products of two operators. One of them depends only on spins 1 and 2 while the other operator depends on spins 3, 4, ... $N$ and equals the product of one-spin operators. Such approach allows us to rewrite the correlation functions as follows:
\begin{eqnarray}\label{Qff1}
&&
p=\frac{1}{2} \tanh  \frac{\beta}{2} \cos ^{N-1} (at),\\
\label{Qff2} &&
q+r=\frac{1}{4}\tanh  ^2 \frac{\beta}{2},\;\; q-r=\frac{1}{4}\tanh  ^2 \frac{\beta}{2}\cos ^{N-2} (2at)\\
\label{Qff3} &&
u=\frac{1}{4}\tanh  \frac{\beta}{2} \cos ^{N-2} (at) \sin (at).
\end{eqnarray}
The correlation functions \eqref{Qff1}-\eqref{Qff3} depend on time periodically and do not decay up to zero at $t\rightarrow \infty$. This is because the coupling constants for all spin pairs are the same in the considered model.

\section{{\small Entanglement of spin pairs in a nanopore}}

In our model elements of the CS matrix of Eq. \eqref{csmatrix} are given by the expressions
\begin{equation}\label{data}
p_1=\frac{1}{4},\; p_2=p_4=\frac{p}{2}, \; p_3 =p_5=-u, \; p_6=q-r, \; p_7 = q+r.
\end{equation}

\begin{figure}
\includegraphics[scale=0.22]{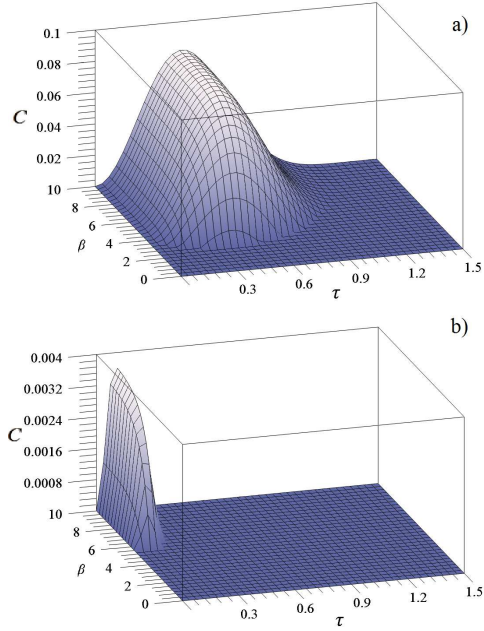}
\caption{Concurrence as a function of the dimensionless time $\tau=at$ and inverse temperatures for different numbers of spins $N=6$ (a), $N=100$ (b).}
\label{pic:graf2}
\end{figure}

Using Eqs. \eqref{llll} one can find that for the system with the density matrix \eqref{redmatr} the parameters $\lambda_1, \lambda_2, \lambda_3, \lambda_4$ needed to calculate the concurrence are 
\begin{equation}\label{eigev} 
\lambda_1=\frac{\sqrt{z}}{2}+w,\; \lambda_2=\frac{\sqrt{z}}{2}-w,\; \lambda_3=\frac{1}{4}-q-r,\; \lambda_4=\frac{1}{4}-q+r,
\end{equation}
where
\begin{equation}\label{zet}
z=\frac{(1+4q)^2-16p^2}{4}, \; w=\sqrt{r^2+4u^2}.
\end{equation}
Using Eqs. \eqref{Qff1} - \eqref{Qff3}, \eqref{eigev}, and \eqref{zet} one can show that $\lambda_1$ is the maximal value in the set $\left\{\lambda_1, \lambda_2, \lambda_3, \lambda_4\right\}$. Then the concurrence (see Eq.\eqref{2qconc})
is
\begin{equation}\label{concur}
C(\rho^{(1,2)})=\max\left\{0, 2(\sqrt{r^2+4u^2}+q)-\frac{1}{2}\right\}.
\end{equation}
The dependence of the concurrence on the inverse temperatures and dimensionless time for different numbers of spins is represented in Fig. \ref{pic:graf2}. It is evident from Fig. \ref{pic:graf2} that the concurrence equals zero at $t=0$. Then the concurrence emerges and its evolution has periodic character. One can see in Fig. \ref{pic:graf2} that the concurrence disappears at shorter times (inside a period $\pi/a$) when the number of the spins increases. In particular, one can obtain from Eq. \eqref{concur}  that the concurrence is absent in the system at $N\rightarrow \infty$.

\section{{\small Quantum discord in spin pairs in a nanopore}}

\begin{figure}
\includegraphics[scale=0.3]{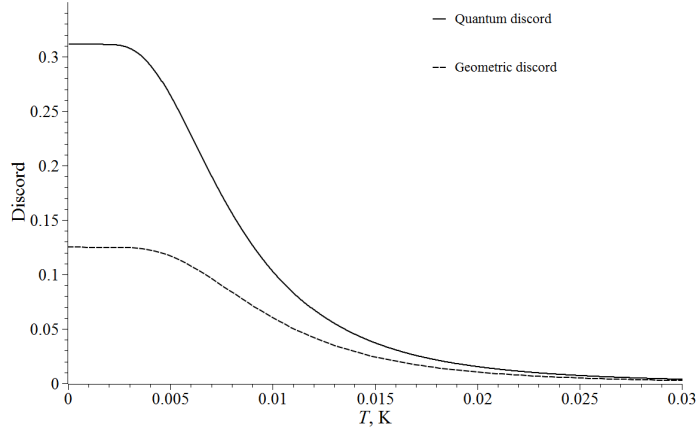}
\caption{The dependences of the quantum and geometric discord for a spin pair in a nanopore on the temperature at $N>>1$, $at\neq \pi l$, $2at\neq \pi n$ ($l,n$ are positive integers). The Larmour frequency  $\omega_0=2\pi\cdot 500\cdot 10^6 s^{-1}$.}
\label{pic:graf3}
\end{figure}

An investigation of the quantum discord in the system described by the density matrix \eqref{redmatr} is a technically complex problem. First we consider the case  $N>>1, \; at\neq \pi l, 2at \neq \pi n$ ($l,\;n$ are positive integers). Then the correlation functions  \eqref{Qff1}-\eqref{Qff3} are getting simple and equal
\begin{equation}
p=u=0, \; q=r=\frac{1}{8} \tanh ^2 \frac{\beta}{2}.
\end{equation}
Thus, the density matrix \eqref{redmatr} has the diagonal Bell form for which the discord $Q$ was evaluated in \cite{luo}. Using the technique \cite{luo}, we find that the discord $Q$ for the spin pair in a nanopore equals
\begin{equation}\label{2qdis}
Q=\frac{1}{4}\left\{(1+8q)\log_2(1+8q)+(1-8q)\log_2(1-8q)\right\}-\frac{1+4q}{2}\log_2(1+4q)-\frac{1-4q}{2}\log_2(1-4q)
\end{equation}
At a low temperature $T\rightarrow 0\; (\beta\rightarrow \infty)$ the asymptotic behavior of the discord is determined as
\begin{equation}\label{discordasymptotic}
Q\approx \frac{3}{4}\log_2 \frac{4}{3}-\frac{\beta e^{-\beta}}{\sqrt{2}}
\end{equation}
Notice that for the temperature limit $T= 0,\; (\beta= \infty)$ we have the expression $Q=\frac{3}{4}\log_2 \frac{4}{3}\cong 0,3113$. 
Surprisingly, the same analitical result was obtained in \cite{datta1} for another special mixed state.
At a high temperature $T\rightarrow \infty \;(\beta \rightarrow 0)$ the asymptotic behavior of the quantum discord is the following
\begin{equation}\label{discordasymptotic_h}
Q\approx \frac{1}{128 \ln 2}\beta^4= \frac{1}{128 \ln 2}\left(\frac{\hbar \omega_0}{k_B T}\right)^4.
\end{equation}
The dependence of the quantum discord on the temperature is represented in Fig. \ref{pic:graf3}. One can see from Fig. \ref{pic:graf3} that  quantum correlations are relatively large only at milli-kelvin temperatures for the Larmour frequency $\omega_0=2\pi\cdot 500 \cdot 10^6s^{-1}$.

The considered model allows us to investigate some peculiarities of the time evolution of the discord. At the time moments $t_l=(1+2l)\pi/(2a),\; l=0,1,2...$ the correlation functions are
\begin{equation}
p=u=0, \; q=\frac{1}{8} \tanh ^2 \frac{\beta}{2}\left[1+(-1)^{N-2}\right],\; r=\frac{1}{8} \tanh ^2 \frac{\beta}{2}\left[1-(-1)^{N-2}\right].
\end{equation}
Essentially that one of the correlation functions, $q$ or $r$, equals zero depending on the parity of the number of spin-carrying molecules in the nanopore. Here the structure of the density matrix coincides with so-called $X$-matrix for which the method of the calculation of the discord was worked out in \cite{ali}. The method \cite{ali} allows us to conclude that the quantum discord equals zero at the time moments $t_l=(1+2l)\pi/(2a)\; (l=0,1,2,...)$. Such is the case because one of the correlation functions equals zero in these time moments. Thus we have ``flickering'' quantum discord which equals zero periodically.

\section{{\small Geometric discord for the $4\times 4$ CS density matrix}}

The geometric discord \cite{dakic} simplifies calculations of the quantum discord.
In particular, it allows us to find the discord for the CS matrix \eqref{csmatrix}. Using the Pauli matrices $\sigma^i_j = 2I_j^i\; (j=1,2; i=x,y,z)$ we can rewrite the density matrix \eqref{csmatrix} in the Bloch form
\begin{equation}\label{bloh}
\begin{array} {r}
\rho\equiv\frac{1}{4}\left[ 1+\sum\limits_{i,j=1}^3 T_{ij}\sigma^i\otimes\sigma^j+  \sum_{i=1}^3 x_i\sigma^i\otimes 1+ \sum_{i=1}^3 y_i 1\otimes\sigma^i \right]\\ \\
=\frac{1}{4}\left[1+(4p_1-1)\sigma^z\otimes\sigma^z +2(p_6+p_7)\sigma^x\otimes\sigma^x+2(p_7-p_6)\sigma^y\otimes\sigma^y -   \right.\\ \\
\left.-4p_3\sigma^z\otimes\sigma^y-4p_5\sigma^y\otimes\sigma^z +4p_4 \sigma^x\otimes 1+4p_2 1\otimes\sigma^x \right],
\end{array}
\end{equation}
where $T$ is the correlation matrix,  equal to
\begin{equation}\label{tensor}
T=\left( \begin{matrix} 2(p_6+p_7) & 0 & 0\\
0 & 2(p_7-p_6)& -4p_5\\
0 & -4p_3 & 4p_1-1
\end{matrix} \right),
\end{equation}
and $\stackrel{\rightarrow}{x}$, $\stackrel{\rightarrow}{y}$ are the Bloch vectors
\begin{equation}
\stackrel{\rightarrow}{x}^T=\left( 4p_4, 0, 0\right),\; \stackrel{\rightarrow}{y}^T=\left( 4p_2, 0, 0\right).
\end{equation}
The geometric discord $Q_g$ for the two-qubit system is given by \cite{dakic}
\begin{equation}\label{geometric_discord}
Q_g=\frac{1}{2}\left(  ||\stackrel{\rightarrow}{x}||^2 + ||T||^2 -k_{max} \right),
\end{equation}
where the spheric (Hilbert-Schmidt) norm $||A||$ is
\begin{equation}
||A||=\sqrt{\texttt{Tr}\left( AA^+ \right)}
\end{equation}
and $k_{max}$ is the maximal eigenvalue of the matrix 
\begin{equation}
K=\stackrel{\rightarrow}{x} \stackrel{\rightarrow}{x}^T+T T^T.
\end{equation}
The geometric discord can be expressed  through $k_1, k_2, k_3$ as
\begin{equation}\label{geometric_discord2}
Q_g=\frac{1}{2}\left(  k_1+k_2+k_3 -k_{max} \right).
\end{equation}
Simple calculations yield the eigenvalues $k_1, k_2, k_3$ of the matrix $K$
\begin{equation}
\begin{array} {c}
k_1=16p_4^2+4(p_6+p_7)^2,\\
k_{2,3}=8p_5^2+2(p_7-p_6)^2+8p_3^2+\frac{1}{2}(4p_1-1)^2\pm\\
\pm \left\{[8p_5^2+2(p_7-p_6)^2-8p_3^2-\frac{1}{2}(4p_1-1)^2]^2+16[p_5(4p_1-1)+2p_3(p_7-p_6)]^2 \right\}^{1/2}.
\end{array}
\end{equation}
In case of a nanopore we can obtain using \eqref{data} the expressions
\begin{equation}
\begin{array} {c}
k_1=4(p^2+4q^2),\\
k_{2,3}=8\left[2u^2+r^2 \pm r\sqrt{r^2+4u^2}\right].
\end{array}
\end{equation}
As a result, the geometric discord has the form
\begin{equation}\label{geometric_discord2}
Q_g=\frac{1}{8}\tanh^4 \frac{\beta}{2}.
\end{equation}
In the limit $T\to0$, the geometric discord $Q_g=1/8=0.125$.
At a high temperature $T\rightarrow \infty, (\beta \rightarrow 0)$ the asymptotic behavior of the geometric discord is the following
\begin{equation}\label{gdiscordasymptotic}
Q_g\approx\frac{1}{128 }\beta^4= \frac{1}{128 }\left(\frac{\hbar \omega_0}{k_B T}\right)^4.
\end{equation}

\section{{\small Conclusion}}

We investigated the entanglement and quantum discord in a nanopore filled with a gas of spin-carrying molecules (atoms). Even when the entanglement is absent the quantum discord can reach large values in the considered system. It reveals significant quantum correlations in the system. They exhibit ``flickering'' character and equal zero periodically in the process of the time evolution of the system.

The density matrix in the considered model belongs to the so-called CS matrices. We evaluated the analytic expression for the concurrence of a general two-qubit CS density matrix. We studied the geometric discord for systems with CS density matrices and obtained an analytic expression for the geometric discord in the considered model.

The model under question confirms once more that entanglement describes only a part of quantum correlations while the discord is a measure of total quantum correlations \cite{vedral_arxiv}.

The authors thank A.I. Zenchuk for useful discussions. The work is supported by the Program of the Presidium of RAS No.8 "Development of methods of obtaining chemical compounds and creation of new materials".

\section*{{\small Appendix. Entanglement of the CS $ 4\times4$ density matrix}}

\setcounter{equation}{0}
\renewcommand{\theequation}{A.\arabic{equation}}

To calculate the entanglement for the general two-qubit density matrix, one should find  the ``spin-flip transformed'' density matrix which is \cite{wootters}
\begin{equation}
\widetilde{\rho}=(\sigma_1^y\otimes \sigma_2^y)\rho^*(\sigma_1^y\otimes \sigma_2^y),
\end{equation}
where $\rho^*$ is the complex conjugate matrix to $\rho$ in the standard basis $\left\{\left|00\right\rangle,\;\left|01\right\rangle,\;\left|10\right\rangle,\;\left|11\right\rangle\right\}$, and $\sigma_y^j$ $(j=1,2)$ is the Pauli matrix. Then it is necessary to calculate the concurrence $C(\rho)$ which determines unambiguously the entanglement $E(\rho)$ \cite{wootters}
\begin{equation}\label{concarrence}
E(\rho)=H\left( \frac{1+\sqrt{1-C^2(\rho)}}{2} \right)
\end{equation}
where $H(x)$ is the Shannon function \cite{shannon}
\begin{equation}\label{shanf}
H(x)=-x\log _2 x -(1-x)\log_2 (1-x).
\end{equation} 
The concurrence is given as \cite{wootters}
\begin{equation}\label{2qconc}\begin{array} {r}
C(\rho)=\max\left\{0,\; 2\lambda-\lambda_1-\lambda_2-\lambda_3-\lambda_4\right\}\\
\lambda=\max\left\{\lambda_1, \lambda_2, \lambda_3, \lambda_4\right\},
\end{array}
\end{equation}
where $\lambda_1,\; \lambda_2,\; \lambda_3,\; \lambda_4$ are the square roots of the eigenvalues of the matrix product $\rho\widetilde{\rho}$.

One can show that the matrix $\widetilde{\rho}$ is a $CS$ matrix if the density matrix $\rho$ is a $CS$ one. It means that the matrix $\rho\widetilde{\rho}$ is also a CS one. Under the orthogonal transformation  
\begin{equation}\label{transformation1}
S=\frac{1}{\sqrt{2}}\left( \begin{matrix} 1 & 0 & 0 & 1 \\
0 & 1 & 1 &0\\
0 & 1 & -1& 0\\
1 & 0 & 0 & -1 \end{matrix} \right)=S^{T}
\end{equation}
any $4\times 4$ CS matrix takes the block-diagonal form consisting of two $2\times2$ subblocks. As a result, the square roots of the eigenvalues of the matrix product $\rho\widetilde{\rho}$ can be written as follows
\begin{equation}\label{l12}
\begin{array} {l}
\lambda_{1,2}=\frac{1}{\sqrt{2}}\left\{{(p_1+p_6)^2-2(p_2+p_4)^2+2(p_3+p_5)^2+(\frac{1}{2}-p_1+p_7)^2}\right. \\ \\
\left. \pm\sqrt{{[(2p_1+p_6-\frac{1}{2}-p_7)^2+4(p_3+p_5)^2][(\frac{1}{2}+p_6+p_7)^2-4(p_2+p_4)^2]}}\right\}^{\frac{1}{2}},
\end{array}
\end{equation}
\begin{equation}\label{l34}
\begin{array} {l}
\lambda_{3,4}=\frac{1}{\sqrt{2}}\left\{{(p_1-p_6)^2-2(p_2-p_4)^2+2(p_3-p_5)^2+(\frac{1}{2}-p_1-p_7)^2}\right.\\ \\
\left. \pm\sqrt{{[(2p_1-p_6-\frac{1}{2}+p_7)^2+4(p_3-p_5)^2][(\frac{1}{2}-p_6-p_7)^2-4(p_2-p_4)^2]}}\right\}^{\frac{1}{2}}.
\end{array}
\end{equation}
Using identities
\begin{equation}\label{ident}
\begin{array} {r}
2\left[(p_1+p_6)^2+ (\frac{1}{2}-p_1+p_7)^2\right]=(2p_1+p_6-\frac{1}{2}-p_7)^2+(\frac{1}{2}+p_6+p_7)^2,\\ \\
2\left[(p_1-p_6)^2+ (\frac{1}{2}-p_1-p_7)^2\right]=(2p_1-p_6-\frac{1}{2}+p_7)^2+(\frac{1}{2}-p_6-p_7)^2,
\end{array}
\end{equation}
we can rewrite finally $\lambda_1, \lambda_2, \lambda_3, \lambda_4$ as
\begin{equation}\label{llll}
\begin{array} {r}
\lambda_1=\frac{1}{2}\left\{\sqrt{(2p_1+p_6-\frac{1}{2}-p_7)^2+4(p_3+p_5)^2}+ \sqrt{(\frac{1}{2}+p_6+p_7)^2-4(p_2+p_4)^2}\right\},\\ \\
\lambda_2=\frac{1}{2}\left|\sqrt{(2p_1+p_6-\frac{1}{2}-p_7)^2+4(p_3+p_5)^2}- \sqrt{(\frac{1}{2}+p_6+p_7)^2-4(p_2+p_4)^2}\right|,\\ \\
\lambda_3=\frac{1}{2}\left\{\sqrt{(2p_1-p_6-\frac{1}{2}+p_7)^2+4(p_3-p_5)^2}+ \sqrt{(\frac{1}{2}-p_6-p_7)^2-4(p_2-p_4)^2}\right\},\\ \\
\lambda_4=\frac{1}{2}\left|\sqrt{(2p_1-p_6-\frac{1}{2}+p_7)^2+4(p_3-p_5)^2}- \sqrt{(\frac{1}{2}-p_6-p_7)^2-4(p_2-p_4)^2}\right|.
\end{array}
\end{equation}
These relationships together with Eqs. \eqref{concarrence}-\eqref{2qconc} give an analytical formula only through the square radicals for calculating the entanglement of formation for the arbitrary two-qubit density matrix \eqref{csmatrix}.

\end{document}